\documentstyle[times,pramana,epsf,floats]{ias}
\begin{document}
\mark{{Gravitational collapse and naked singularities}{Tomohiro Harada}}
\title{Gravitational collapse and naked singularities}

\author{Tomohiro Harada}
\address{Astronomy Unit, School of Mathematical Sciences,
Queen Mary, University of London, Mile End Road,
London E1 4NS, UK}
\keywords{general relativity, spacetime singularity}
\pacs{04.20.Dw}
\abstract{
Gravitational collapse is one of the most striking phenomena
in gravitational physics. The cosmic censorship conjecture
has provided strong motivation for research in this field. 
In the absence of general proof for censorship, many 
examples have been proposed, in which naked singularity 
is the outcome of gravitational collapse. 
Recent development has revealed that 
there are examples of naked singularity formation
in the collapse of physically reasonable matter fields,
although the stability of these examples is still uncertain. 
We propose the concept of `effective naked singularities',
which will be quite helpful because general relativity has 
the limitation of its application for high-energy end. 
The appearance of naked singularities is not detestable but 
can open a window for the new physics of strongly curved spacetimes.
}
\maketitle
\section{Introduction}
If a pressure gradient force is not sufficiently strong,
a body can continue collapsing due to its self-gravity.
This phenomenon is called gravitational collapse.
The astrophysical relevance of gravitational collapse
is now robust.
We have observational evidence for 
massive and/or supermassive black holes.
It is well understood that there exists an upper limit 
to the maximum possible mass of a spherical body 
of cold nuclear matter.
We can also naturally consider that 
black holes may have formed from cosmological perturbations
in the early stages of universe.

The singularity theorems (see, e.g.,~\cite{he1973}) state that there exist spacetime 
singularities in generic gravitational collapse.
At singularities the smoothness of the spacetime
metric is lost. More precisely, a singularity is not 
regarded as a point in the spacetime manifold but one in the boundary of
spacetime (see~\cite{clarke1993}). Classical physics 
cannot be applied to spacetime
singularities because classical physics implicitly 
assumes the smoothness of spacetime,
where and hereafter I refer to all physics which do not contain
quantum gravity as classical physics.
This implies the limitation of classical physics.

In spite of this limitation, it is possible 
to have some kind of future predictability using only classical 
physics if all singularities are hidden from 
physically relevant regions. 
Penrose~\cite{penrose1969} conjectured that a system
which evolves, according to classical general relativity
with reasonable equations of state,
from generic non-singular initial data
on a suitable Cauchy hypersurface,
does not develop any spacetime singularity 
which is visible from infinity.
This conjecture is actually a weak version of censorship.
If this conjecture is true, we can predict the whole
evolution using classical physics for 
the spacetime region outside black holes.
The strong version~\cite{penrose1979} states that a physically reasonable 
classical spacetime is globally hyperbolic.
This conjecture claims that classical physics 
can predict the evolution of the whole spacetime.
A singularity which is censored by the strong version 
of censorship is called a naked singularity.
A naked singularity which is censored by the weak version
is called a globally naked singularity.
A naked singularity which is not a globally naked singularity
is called a locally naked singularity.

The cosmic censorship has not been proved yet in spite of a huge 
amount of effort. It is clear that there is ambiguity in
the definition of physical reasonableness. 
We can consider two aspects of physical reasonableness,
one for matter fields and the other for initial data.
For matter fields, many people have considered that energy conditions
should be satisfied. Furthermore, we can consider more concrete models,
such as a perfect fluid with reasonable equations of state,
elementary fields, and so on.
For initial data, we should restrict our attention to non-singular 
initial data. Moreover, we can consider sufficiently smooth 
initial data. The initial data should 
be generic in some appropriate topology.  

It is true that the cosmic censorship conjecture has long provided 
researchers with strong motivation into studies of gravitational collapse
but I think now that we should regard it as one of aspects in 
gravitational collapse physics. I would like to come back to the 
original standpoint. How is physical gravitational collapse?
How do black holes form?
How do spacetime singularities form?
 
\section{Spherically symmetric dust collapse}
The spherically symmetric collapse of a dust fluid 
is the simplest nontrivial example of gravitational collapse.
Dust means a fluid with vanishing pressure.
This model is pedagogical to understand how the situation gets
complicated if we include inhomogeneity in the model.

\subsection{Homogeneous dust ball}
\begin{figure}[htbp]
\epsfysize=4cm
\centerline{\epsfbox{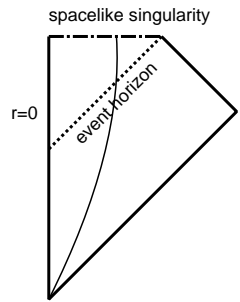}}
\caption{Spacetime diagram of the collapse of a homogeneous model.}
\label{fig:os}
\end{figure}

In the spherical collapse of a homogeneous dust ball, 
the dust interior and vacuum exterior are described by the 
collapsing Friedmann solution 
and the Schwarzschild solution, respectively.
This collapse solution, which is called the Oppenheimer-Snyder 
solution~\cite{os1939},
has given a picture of a black hole as a final state of gravitational
collapse. The mass $M$ of the dust ball begins to collapse and  
the radius of the dust ball gets smaller and smaller.
At some moment it equals the Schwarzschild radius $2M$, which 
corresponds to the formation of an event horizon.
During this process, the redshift of the light ray which is emitted
at the surface increases and 
becomes infinite at the formation of an event horizon.
Subsequently, a singularity forms everywhere at the same time
in the dust interior.
The singularity is spacelike and completely hidden by the horizon.
No causal curve can emanate from the singularity.
The causal structure of this spacetime is depicted
in Fig.~\ref{fig:os}.
This picture is generally accepted
as a final fate of gravitational collapse in realistic situations.

\subsection{Inhomogeneous dust ball}
\begin{figure}[htbp]
\epsfxsize=8cm
\centerline{\epsfbox{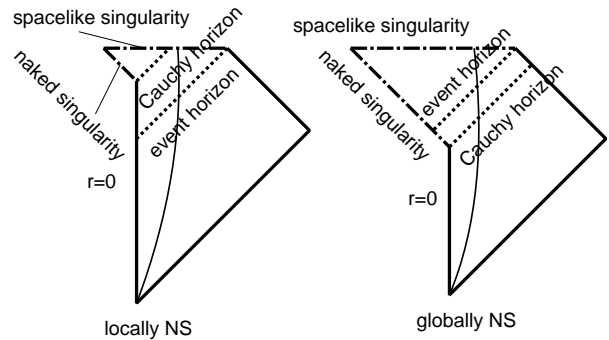}}
\caption{Spacetime diagram of the collapse of a inhomogeneous model.}
\label{fig:ltb}
\end{figure}

The situation is completely different if we allow inhomogeneity
in the model.
The collapse of an inhomogeneous dust ball is described by an exact solution,
the Lema\^{i}tre-Tolman-Bondi (LTB) 
solution~\cite{ltb1933}.
This solution is general in a sense that it has two 
arbitrary functions, which correspond to 
initial density and velocity profiles.
In this solution, a singularity forms at the centre, which is 
called a shell-focusing singularity.
The shell-focusing singularity can be a naked singularity.
This was first found numerically~\cite{es1979} 
and subsequently proved mathematically~\cite{christodoulou1984}.
The conditions for appearance of naked singularities have 
been extensively investigated~\cite{jd1993sj1996jj1997}.
The result is that naked singularities
are the generic outcome of this collapse models which develop
both from non-singular initial data and from sufficiently smooth 
initial data. The naked singularity is globally naked singularities
or locally naked singularities, depending on the choice of 
the arbitrary functions. 
The causal structure is depicted in Fig.~\ref{fig:ltb}.

For a while we restrict our attention to marginally bound 
collapse with initial data where
the initial density profile is given by an analytic function with 
respect to the locally 
Cartesian coordinates. In this case, the structure of 
naked shell-focusing singularity has been revealed.
The redshift of the first null ray which emanates from the singularity
is finite but 
becomes infinite immediately for later null rays~\cite{christodoulou1984}.
This naked singularity is curvature strong~\cite{newman1986,djd1999}.
From this singularity, not only radial null rays but also 
nonradial null rays emanate, which implies that the 
singularity is not point-like~\cite{mn2001,djd2002}.

From this structure of the naked shell-focusing 
singularities, it is undoubted that they are genuine
and inextendible singularities. 
On the other hand, the dust fluid would not be physically 
reasonable because vanishing pressure should not be physically 
accepted. The recent development in gravitational collapse physics
has provided modern examples of naked singularities where 
matter fields are sufficiently physically reasonable.
\section{Modern examples of naked singularities}
The problem of gravitational collapse involves
dynamical and inhomogeneous spacetimes in its nature.
It is difficult to obtain somewhat general exact solutions
in this situation except for in limited systems such as
the LTB solutions. Hence, it is natural to 
try to understand the qualitative nature of gravitational 
collapse by numerical simulations.
This approach is getting more and more popular in the progress 
of numerical relativity and has a potential
to discover new unexpected phenomena in general relativity.
One of the most successful ones will be critical behaviour
at the black hole threshold.

\subsection{Black hole threshold}
The critical behaviour in gravitational collapse was discovered
in a spherical system of a massless scalar field~\cite{choptuik1993},
subsequently has been found in a variety of systems, such as 
axisymmetric gravitational waves~\cite{ae1993}, 
spherical system of a radiation fluid~\cite{ec1994}, 
spherical system of a perfect fluid with the equation of 
state $P=k\rho$~\cite{nc2000}, and so on.
Suppose we have a one-parameter family of initial data sets
parametrised by a parameter $p$, 
which evolve through the Einstein equations,
and the system has two extremal end states,
one is a collapse to a black hole and the other is a dispersion 
to infinity. For instance, the data with a sufficiently large value
for $p$ evolves to a black holes and the one with a sufficiently 
small value for $p$ evolves to a dispersion.
Then, there should be a critical value $p^{*}$ at the threshold 
between two end states.
For a near-critical case $p\approx p^{*}$, 
the collapse first approaches a self-similar 
solution, which is called a critical solution, and then 
it deviates away from the critical solution.
For a supercritical and near-critical case, a black hole 
forms finally and
the mass $M_{BH}$ of the formed black hole satisfies the 
scaling law $M_{BH}\propto |p-p^{*}|^{\gamma}$,
where $\gamma$ is a positive constant called 
a critical exponent. These phenomena are universal to 
the choice of the one-parameter family of initial data sets.

This phenomenon is well described by the intermediate 
behaviour around the intermediate attractor~\cite{kha1995}.
The critical solution is identified with a self-similar 
solution with a single unstable mode. 
Here we consider a space of functions of $x$,
where we put $x=\ln [r/(-t)]$ and $\tau=-\ln(-t)$
for some appropriate time and radial coordinates, $t$ and $r$. 
Then a self-similar solution corresponds to a fixed point,
while a non self-similar solution moves along a curve 
in this space as $\tau$ increases. 
The critical solution is a fixed point 
with a single unstable direction. In other words, 
this fixed point has a stable manifold of codimension one,
which is called a critical surface.
A one-parameter family of initial data sets,
which correspond to a curve in this space, 
generically has an intersection 
with the critical surface. Since the intersection evolves to 
the critical solution, the intersection corresponds to the 
data with the critical value $p^{*}$.
For a near critical case, $p\approx p^{*}$, the initial data near the 
intersection first shadows the critical collapse but deviates
away later because of the growth of the unstable mode.
It can be shown that the critical
exponent $\gamma$ is equal to the inverse of the eigenvalue 
of the unstable mode. The value calculated from the eigenvalue 
analysis shows a good agreement with the value obtained 
from the results of numerical simulations.

\begin{figure}[htbp]
\epsfysize=4cm
\centerline{\epsfbox{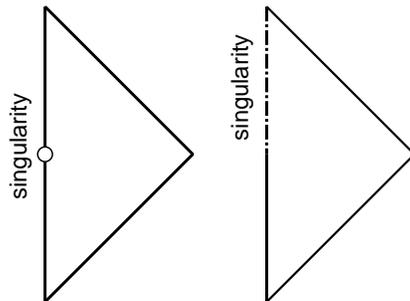}}
\caption{Spacetime diagram of the 
critical solution found in the spherical collapse 
of a massless scalar field.}
\label{fig:penrose_choptuik}
\end{figure}

For the supercritical collapse, the formed black hole mass $M_{BH}$ is 
proportional to $|p-p^{*}|^{\gamma}$. From a dimensional consideration,
we can derive that the curvature strength scales as $M_{BH}^{~~-2}$
in the region just outside the event horizon. Then, 
let us consider the limit to the critical collapse from the 
supercritical collapse, i.e., $p\to p^{*}$.
In this limit, the black hole mass approaches zero and 
the curvature just outside the event horizon is going 
to infinity. Since we have arbitrarily strong curvature 
outside the event horizon by fine-tuning,
the black hole threshold can be regarded as a naked singularity.
This intuitive consideration is confirmed by more straightforward
work~\cite{mg2003}, in which the causal structure of 
the critical solution found in the spherical system of a 
massless scalar field was shown to be naked-singular and 
given by one of the two possible spacetime diagrams 
depicted in Fig.~\ref{fig:penrose_choptuik}.

It is now clear that critical collapse at the 
black hole threshold provides examples of naked singularities
which form in the collapse of physically reasonable matter 
fields with non-singular initial data.
However, it is also clear that the critical collapse is unstable and 
realised only as a result of exact fine-tuning.
This implies that the formation of these naked singularities 
is not generic.

\subsection{Self-similar attractor}
As we have seen, critical behaviour in gravitational collapse
is associated with a self-similar solution with a single unstable 
mode. It is expected that if we have a self-similar solution
with no unstable mode, it will act as an attractor.

In fact, this is the case for a perfect fluid with the 
equation of state $P=k\rho$ for $0<k\le 0.03$.
In the result of numerical simulations~\cite{hm2001}, it was 
observed that 
a spherically collapsing perfect fluid approaches a self-similar 
solution from many initial data sets 
without any fine-tuning of initial data parameters.
The approached solution is different from the critical solution
and turns out to be one of the discrete set of 
self-similar solutions,
which was previously discovered numerically~\cite{op198790}.
This solution is called the general relativistic Larson-Penston
solution. This solution is stable against spherical linear perturbations,
while other self-similar solutions are unstable against 
those perturbations~\cite{hm2001,harada2001}.
These results strongly suggest that the general relativistic 
Larson-Penston self-similar solution acts as an attractor
in the gravitational collapse of a perfect fluid.

\begin{figure}[htbp]
\epsfysize=4cm
\centerline{\epsfbox{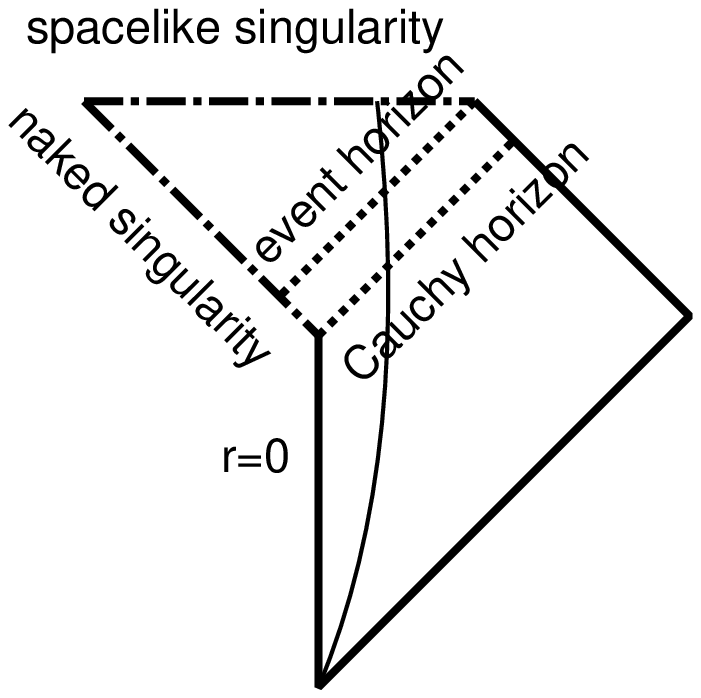}}
\caption{Spacetime diagram of the general relativistic 
Larson-Penston solution.}
\label{fig:op}
\end{figure}

The general relativistic Larson-Penston solution is 
interesting in the context of cosmic censorship~\cite{op198790}.
This solution describes the formation of naked singularity 
in the collapse of a perfect fluid from sufficiently smooth 
initial condition for $0<k\alt 0.0105$. This singularity 
can be globally naked after an appropriate matching to the 
Schwarzschild solution as is depicted in Fig.~\ref{fig:op}. 
For $k\agt 0.0105$, this singularity is spacelike.

Since the general relativistic Larson-Penston solution 
is an attractor solution, the naked singularity
formation is the outcome of generic gravitational 
collapse of a perfect fluid with the equation of state 
$P=k\rho$ ($0<k\alt 0.0105$) in spherical symmetry.
No fine-tuning is needed to realise the naked singularity unlike
in the black hole threshold.
This is also consistent with the result of numerical 
simulations~\cite{harada1998}.
As a result, if we can restrict our attention to 
spherically symmetric collapse, 
we can conclude generic violation of cosmic censorship.

\begin{figure}[htbp]
\epsfysize=4cm
\centerline{\epsfbox{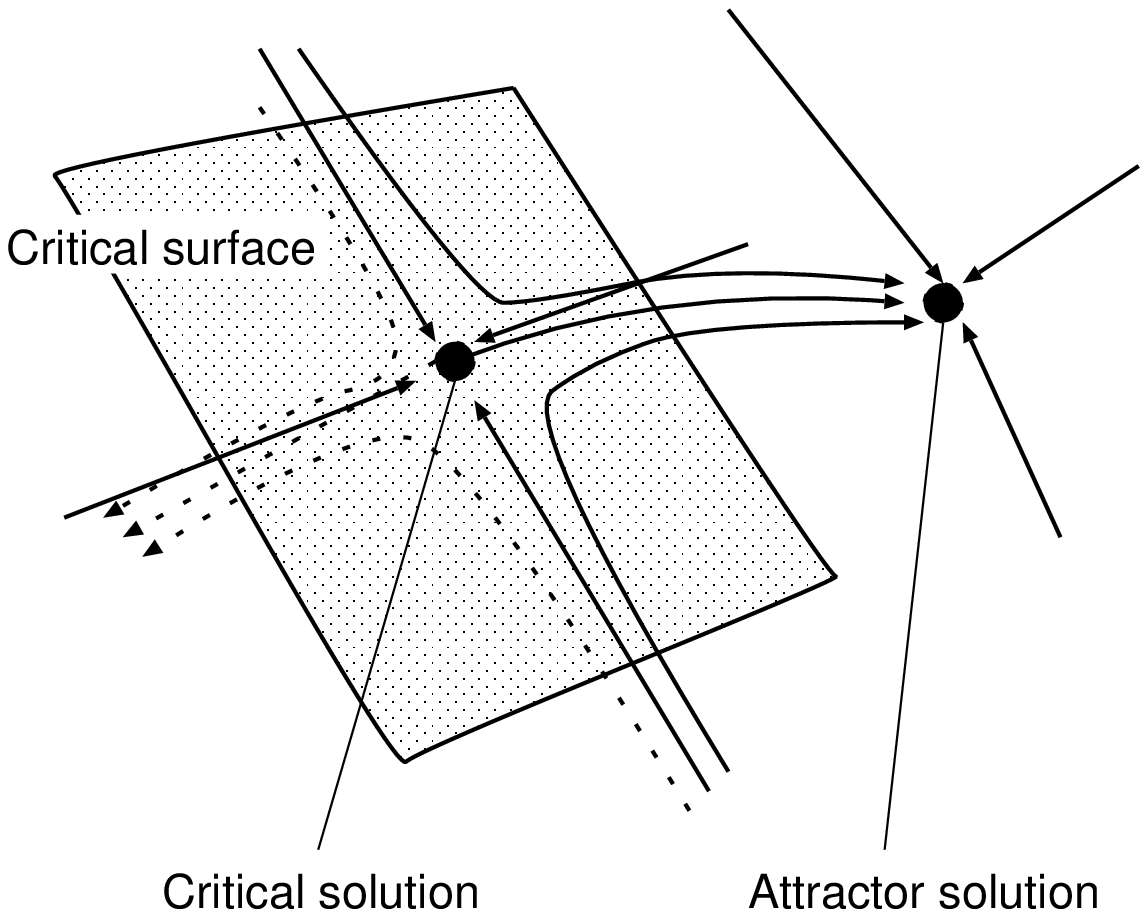}}
\caption{Schematic figure of qualitative properties of 
dynamical evolution of spacetimes  
with a self-similar attractor and a 
self-similar critical solution.} 
\label{fig:flow}
\end{figure}

This example is also interesting in the context of 
dynamical properties of 
spacetimes in collapsing situations.
This is actually the first demonstration of an attractor self-similar 
solution in generic spherical collapse in general relativity.
The critical behaviour and self-similar attractor in 
spherical gravitational collapse suggest the validity of 
the so-called self-similarity hypothesis~\cite{cc1999},
i.e., the expectation that self-similar solutions can describe 
the asymptotic or intermediate behaviour of more
general solutions in a variety of systems.
For example, when we have both a self-similar attractor and 
a self-similar critical solution, which is the case
for the spherical system of a perfect fluid with 
the equation of state $P=k\rho$ ($0<k\le 0.03$), the 
qualitative properties of spacetime evolution can be depicted
as in Fig.~\ref{fig:flow}.
The self-similarity hypothesis gives an important and useful 
view point for the general features of dynamical 
evolution of spacetimes through 
Einstein's field equations.

\subsection{Highly nonspherical collapse}
There are many other explicit examples of naked singularities, 
most of which are spherically symmetric.
Although some of them provide an interesting point of view 
for gravitational collapse physics,
unfortunately I cannot mention all of them because of 
space limitation.
Here I would like to comment only on highly nonspherical collapse.

Among highly nonspherical collapse systems, the simplest one
is cylindrically symmetric collapse.
This system is described in time and one-dimensional space, 
which is the same as the 
spherically symmetric system. On the other hand,
this system has one degree of freedom for gravitational waves,
while the spherically symmetric system has none.
The cylindrically symmetric system is the simplest system
which contains both a collapsing body and gravitational waves.
It has been proved that no horizon exists in 
this system~\cite{thorne1972,hayward2000}.
This implies that if the cylindrical collapse ends in
singularity, it must be naked.
Cylindrically symmetric collapse models 
with and without (counter)rotation
have been investigated by several 
authors~\cite{at1992,echeverria1993,chiba1996,pw2000,nolan2002}.
A strong gravitational wave burst just before the singularity formation 
was reported~\cite{echeverria1993}. 
I think many things are not fixed yet in the study 
of cylindrically symmetric collapse. 
I expect significant progress in the near future.

For axisymmetric case, we have a much less knowledge.
The formation of singularity without 
apparent horizons has been reported numerically,
which is called a spindle singularity~\cite{nmms1982,st199192}.
However, it is well understood that a singularity 
without apparent horizon in some choice of time slicings
does not necessarily mean a genuine naked singularity.
It is still an open problem whether the spindle singularity
reported by the result of numerical simulations is naked or not.

For completely general case, very little is known.
Nevertheless we have a hoop conjecture~\cite{thorne1972}. 
The conjecture claims that black holes 
with horizons form when and only when a mass $M$ gets compacted
into a region whose circumference in every direction is 
$\alt 4\pi M$. It is interesting to note that 
this conjecture does not 
assume the cosmic censorship.
There is no proof for or no counterexample
against this conjecture. Recent progress in this conjecture
has been made in higher dimensional 
gravity.

For highly nonspherical case, numerical relativity will 
give us a better insight in particular for cylindrically 
symmetric and axisymmetric cases. For completely general 
case, we will need extremely high performance computer 
because very high resolution will be necessary to 
solve the problem of spacetime singularities formed 
in gravitational collapse.
\section{Physics around naked singularities}
Since regions around naked singularities 
are regular, we can apply classical physics 
to these regions. Within the domain of dependence 
we can predict the evolution using classical physics.
In principle we can expect something 
observable as a sign of extremely high curvature
exposed to an observer.
Here we see two examples, the evolution of 
nonspherical linear perturbation
and the effect of quantum field theory in 
curved spacetime during the formation of naked singularities.

\subsection{Nonspherical perturbations and gravitational waves}
\begin{figure}[htbp]
\epsfysize=3cm
\centerline{\epsfbox{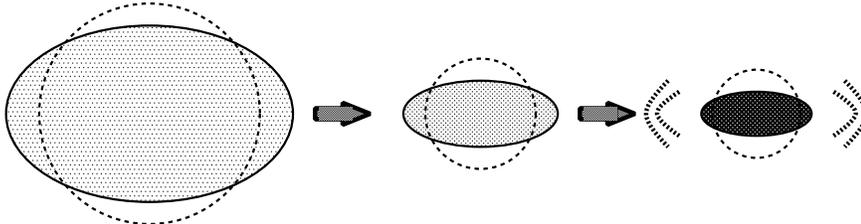}}
\caption{Schematic figure of nonspherical perturbations 
in spherical collapse background.}
\label{fig:perturbation}
\end{figure}

Most known examples of naked singularities formed 
in gravitational collapse
are spherically symmetric. It is interesting to see whether 
these examples are stable or not against nonspherical perturbation.
Nonspherical perturbations involve gravitational waves 
as seen in Fig.~\ref{fig:perturbation}.
Nonspherical metric and matter perturbations on 
spherically symmetric spacetimes
are expanded by scalar, vector and tensor harmonics
with quantum numbers $l$ and $m$.
If we linearise these perturbations, 
the partial differential equations
for gauge invariant perturbation quantities are 
obtained from Einstein's field equations and 
these perturbations are decomposed into two sectors
with their parity.

When we consider nonspherical perturbations of the LTB spacetime,
the field equations for the perturbations result in a set of partial
differential equations, among which main equations are 
wave equations in the LTB spacetime background.
These equations are well integrated by numerical simulations
up to near the Cauchy horizon.
The result for $l=2$ is the following~\cite{inh19989900}.
There is no significant growth of the energy flux of gravitational 
waves.
On the other hand, some tetrad components of the Weyl curvature 
are diverging in an approach to the Cauchy horizon.
The reason why this is possible is that the energy flux of gravitational
waves comes from the first-order time derivative of the metric 
perturbation while the Weyl curvature is proportional to the 
second-order time derivative. 
The metric perturbations near the Cauchy horizon
are described by the combination of the regular term and 
the term proportional to $(t_{CH}-t)^{\alpha}$, where $t_{CH}$
is the time of the Cauchy horizon appearance and $1<\alpha <2 $.
The diverging Weyl curvature will imply some 
instability of the Cauchy horizon.

The pure rotation of the body belongs to the odd-parity perturbation
for $l=1$. In linear order analysis, we have no significant 
growth of perturbation on the Cauchy horizon. 
The centrifugal force due to rotation, which is essential
for the collapse of a homogeneous dust ball, is the second-order
effect. Even for $l=2$ perturbations, it is not clear how
the diverging Weyl curvature in linear order analysis affects
the Cauchy horizon.
These considerations strongly suggest the importance of nonlinear 
analysis on nonspherical perturbations. 

%%%%%%%%%%%%%%%%%%%%%%%%%%%%%%
This is one of the first systematic studies on
nonspherical perturbations of spherically symmetric 
naked-singular spacetimes.
%%%%%%%%%%%%%%%%%%%%%%%%%%%%%%
It will be very interesting 
to study nonspherical perturbations of other examples
of naked singularities or to adopt more general approach
on the stability of naked singularity in 
spherically symmetric collapse
against nonspherical perturbations.
%%%%%%%%%%%%%%%%%%%%%%%%%%%%%%
Important results have been obtained for the 
stability of critical solutions against nonspherical 
perturbations~\cite{mg1999,gundlach2002}.
%%%%%%%%%%%%%%%%%%%%%%%%%%%%%%
\subsection{Quantum field theory in curved spacetime}
When quantum field theory is applied to gravitational 
collapse to a black hole,
the constant thermal radiation, which is called the 
Hawking radiation, is derived.
For naked-singular spacetimes, the extremely high-curvature
region dynamically forms outside horizons.
This fact suggests explosive particle creation 
around the forming naked singularity.

To estimate quantum particle creation,
the Bogoliubov coefficients are calculated.
For this purpose,
we have to obtain mode functions in curved spacetimes.
If we adopt the geometric optics approximation,
this problem reduces to determining the `double-null 
map' $v=G(u)$ between 
Eddington-Finkelstein
outgoing and ingoing null coordinate $u$ and $v$
at the asymptotic region 
for outgoing and ingoing null rays, respectively, where 
the former is a reflection of the latter
at the symmetric centre.
This double-null map is obtained analytically for the   
self-similar LTB spacetime with the Schwarzschild 
exterior~\cite{bsvw1998,vw1998}.
The resultant energy flux due to the particle emission
is diverging in an approach to the Cauchy horizon.
The spectrum of the emission is different from thermal radiation.

\begin{figure}[htbp]
\epsfysize=4cm
\centerline{\epsfbox{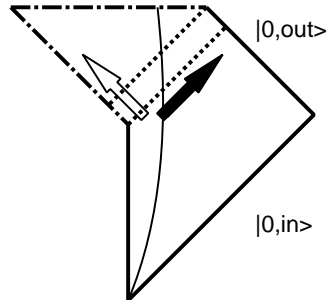}}
\caption{Schematic figure of quantum particle emission
from a forming naked singularity. There are an outgoing 
positively diverging energy flux
and an ingoing negatively diverging energy 
flux near the singularity.}
\label{fig:sc}
\end{figure}

For two-dimensional case, the geometrical optics approximation is exact
and an unambiguous expression is available for the quantum stress-energy 
tensor. The two-dimensional analogue of the four-dimensional LTB spacetime
is obtained by removing the angular sector from the metric tensor.
Then the physical mechanism of this particle emission can be discussed
using the expectation value of the stress-energy tensor of quantum 
fields~\cite{ih2001}.
In this model, the positive energy gradually gathers around
the centre during the collapse, leaving the negative-energy region 
around the positive-energy central region. Just before the 
singularity formation, the accumulated positive energy 
around the centre is emitted as an outgoing flux at a burst.
Schematically, Fig.~\ref{fig:sc} shows the energy flow 
around the naked singularity.

Although the self-similar LTB spacetime is simple and 
useful, it is somewhat unphysical because 
it does not admit Taylor expandable initial 
density profile at the centre. For the LTB spacetimes 
which admit Taylor expandable initial density profile,
it is difficult to obtain the full analytic expression of 
the double-null map. Then, the double-null map was
obtained numerically~\cite{hin2000} and its qualitative property
was also indicated analytically~\cite{ts2001}.
The resultant energy flux is also diverging in an approach
to the Cauchy horizon.

The diverging energy flux along the Cauchy horizon 
suggests a possibility that the Cauchy horizon is 
unstable due to quantum effects and 
could be altered if the quantum gravity effects are
seriously taken into account.
However, there should be two major objections 
about these calculations.
One objection is that the geometrical optics approximation
has not been justified yet for the four dimensional LTB spacetimes.
The difficulty come from the fact that quanta go through extremely
curved region for naked singularity formation unlike 
for black hole formation.
The other is that the amount of energy flux 
due to the particle creation is highly subjected to the 
cutoff scale if we introduce it.
If we set the cutoff scale to be the Planck scale,
the total radiated energy does not 
exceed the Planck energy~\cite{hinstv2001}.
Although we could extrapolate this perturbative calculation
up to arbitrarily high energy scale, 
the calculated result has no justification if we do not have knowledge 
on physics beyond the cutoff scale.
Because of these objections, I have to say that 
the divergent energy flux calculated in the LTB spacetimes
is not sufficiently robust yet,
even though it may be suggestive.
\subsection{Effective naked singularities}
Finally, I would like to propose the concept of `effective naked
singularities'.
%%%%
This has been coming up
from the interaction of people in this field 
rather than is my own invention.
For example, similar discussions have been made 
in Refs.~\cite{joshi1993,jhingan1998}. 
See Ref~\cite{hn2004} for the mathematical definition
of this concept as the 'border of spacetime'. 
%%%%

For naked singularities, 
the curvature strength outside the horizons is typically diverging
in an approach to the singularity.
The cosmic censorship claims that naked singularities
are not generic outcome of gravitational collapse 
of physically reasonable matter fields in classical 
general relativity.

\begin{figure}[htbp]
\epsfysize=4cm
\centerline{\epsfbox{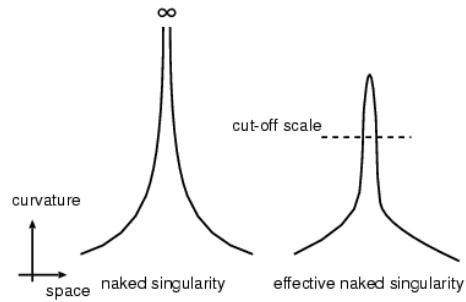}}
\caption{Schematic figure of effective naked singularities.}
\label{fig:effective_ns}
\end{figure}

However, it is now reasonable to consider that 
classical general relativity has 
a limitation towards the high energy end.
We can consider a cutoff energy scale beyond which 
classical general relativity gives no longer
good approximation and quantum gravity should be 
needed.
Then it is natural to modify the classical notion
of naked singularities.
Let us consider that a spacetime region outside horizons.
In this region, we can measure the energy scale of 
the spacetime by the curvature strength even if 
the spacetime is empty.
If this energy scale is beyond the cutoff scale of 
classical general relativity, 
classical general relativity is not valid there and 
the spacetime should be modified considerably 
due to the quantum gravity effects.
I will call this region an effective naked singularity.
Figure \ref{fig:effective_ns} gives an illustration
of an effective naked singularity.

As we have seen, we have already examples of naked singularities
which form in the collapse of physically reasonable matter fields,
although they may not be stable.
The important implication of effective naked singularities 
is that they appear with nonzero probability in the collapse
of physically reasonable matter fields.
This is because we can realise the appearance of effective 
naked singularities if we prepare an initial data set which is 
sufficiently close to the one corresponding to naked singularity
appearance.
Since effective naked singularities are outside horizons,
we can no longer predict our future without quantum gravity.
In this sense, we can claim that the cosmic censorship conjecture 
is effectively violated, i.e., classical physics which does not 
include quantum gravity has a limitation in its 
future predictability with nonzero probability.
%%%%%%%%%%%%%%%%%%%%%%%%%%%%
%The concept of effective naked singularities has been 
%mathematically defined
%as the `border of spacetime' in Ref.~\cite{hn2004}.
%%%%%%%%%%%%%%%%%%%%%%%%%%%%
\section{Summary}
We have now a large number of examples of naked singularities.
Among them, we can regard the black hole threshold and 
the self-similar attractor
as naked singularities formed in the collapse of physically 
reasonable matter fields.
The black hole threshold is unstable and the stability of 
the self-similar attractor against nonspherical perturbations
is not clear.
Therefore, it is still uncertain whether or not naked singularities
are censored in physically reasonable gravitational collapse.
However, these modern examples of naked singularities strongly 
suggest that effective naked singularities appear 
with nonzero (maybe not too small) probability for physically 
reasonable matter fields.
Therefore, naked singularities
are within the scope of physics.
Naked singularities are worth studying in classical physics and
in some form of quantum gravity,
as the appearance of extremely strong curvature is 
in principle observable.

\end{document}